\documentclass{aastex631}
\usepackage{amsmath}
\usepackage{savesym}
\savesymbol{tablenum}
\usepackage{siunitx}
\restoresymbol{SIX}{tablenum}

\submitjournal{ApjL}
\shorttitle{WASP-189b Transmission spectrum}
\shortauthors{Sreejith et al.}

\begin{document}
\title{CUTE reveals escaping metals in the upper atmosphere of the ultra-hot Jupiter WASP-189b}


\correspondingauthor{A. G. Sreejith}
\email{sreejith.aickara@lasp.colorado.edu, sreeijth.aickara@oeaw.ac.at}
\author[0000-0002-4166-4263]{A. G. Sreejith}
\affiliation{Laboratory for Atmospheric and Space Physics, University of Colorado, UCB 600, Boulder, CO, 80309, USA}
\affiliation{Space Research Institute, Austrian Academy of Sciences, Schmiedlstrasse 6, 8042 Graz, Austria}
\author[0000-0002-1002-3674]{Kevin France}
\affiliation{Laboratory for Atmospheric and Space Physics, University of Colorado, UCB 600, Boulder, CO, 80309, USA}
\author[0000-0003-4426-9530]{Luca Fossati}
\affiliation{Space Research Institute, Austrian Academy of Sciences, Schmiedlstrasse 6, 8042 Graz, Austria}

\author[0000-0003-3071-8358]{Tommi T. Koskinen}
\affiliation{Lunar and Planetary Laboratory, University of Arizona, Tucson, AZ 85721--0092}
\author[0000-0002-4701-8916]{Arika Egan}
\affiliation{Laboratory for Atmospheric and Space Physics, University of Colorado, UCB 600, Boulder, CO, 80309, USA}

\author[0000-0001-9207-0564]{P. Wilson Cauley}
\affiliation{Laboratory for Atmospheric and Space Physics, University of Colorado, UCB 600, Boulder, CO, 80309, USA}

\author{Patricio. E. Cubillos}
\affiliation{INAF – Osservatorio Astrofisico di Torino, Via Osservatorio 20, 10025 Pino Torinese, Italy}

\affiliation{Space Research Institute, Austrian Academy of Sciences, Schmiedlstrasse 6, 8042 Graz, Austria}
\author[0000-0002-0506-0825]{S. Ambily}
\affiliation{Laboratory for Atmospheric and Space Physics, University of Colorado, UCB 600, Boulder, CO, 80309, USA}
\author{Chenliang Huang}
\affiliation{Shanghai Astronomical Observatory, CAS, Shanghai, 200030, China}
\author{Panayotis Lavvas}
\affiliation{Groupe de Spectroscopie Mol\'eculaire et Atmosph\'erique, Universit\'e de Reims, Champagne-Ardenne, CNRS UMR-7331, France}
\author[0000-0002-2129-0292]{Brian T. Fleming}
\affiliation{Laboratory for Atmospheric and Space Physics, University of Colorado, UCB 600, Boulder, CO, 80309, USA}
\author[0000-0002-0875-8401]{Jean-Michel Desert}
\affiliation{Anton Pannekoek Institute for Astronomy, University of Amsterdam, P.O. Box 94249, Noord Holland, NL-1090GE Amsterdam, the Netherlands}
\author[0000-0001-7131-7978]{Nicholas Nell}
\affiliation{Laboratory for Atmospheric and Space Physics, University of Colorado, UCB 600, Boulder, CO, 80309, USA}
\author[0000-0001-7624-9222]{Pascal Petit}
\affiliation{Institut de Recherche en Astrophysique et Plan\'etologie, Universit\'e de Toulouse, CNRS, CNES, 14 avenue Edouard Belin, 31400 Toulouse, France}
\author[0000-0001-5371-2675]{Aline Vidotto}
\affiliation{Leiden Observatory, Leiden University, PO Box 9513, 2300 RA, Leiden, the Netherlands}

\begin{abstract}
Ultraviolet observations of Ultra-hot Jupiters (UHJs), exoplanets with temperatures over 2000\,K, provide us with an opportunity to investigate if and how atmospheric escape shapes their upper atmosphere. Near-ultraviolet transit spectroscopy offers a unique tool to study this process owing to the presence of strong metal lines and a bright photospheric continuum as the light source against which the absorbing gas is observed. WASP-189b is one of the hottest planets discovered to date, with a day-side temperature of about 3400\,K orbiting a bright A-type star. We present the first near-ultraviolet observations of WASP-189b, acquired with the Colorado Ultraviolet Transit Experiment ($CUTE$). $CUTE$ is a 6U NASA-funded ultraviolet spectroscopy mission, dedicated to monitoring short-period transiting planets. WASP-189b was one of the $CUTE$ early science targets and was observed during three consecutive transits in March 2022. We present an analysis of the $CUTE$ observations and results demonstrating near-ultraviolet (2500--3300~\AA) broadband transit depth ($1.08^{+0.08}_{-0.08}\%$) of about twice the visual transit depth indicating that the planet has an extended, hot upper atmosphere with a temperature of about 15000\,K and a moderate mass loss rate of about \SI{4e8}{\kg\per\second}. We observe absorption by Mg{\sc ii} lines ($R_p/R_s$ of $0.212^{+0.038}_{-0.061}$) beyond the Roche lobe at $>$4$\sigma$ significance in the transmission spectrum at a resolution of 10~\AA, while at lower resolution (100~\AA), we observe a quasi-continuous absorption signal consistent with a ``forest" of low-ionization metal absorption dominated by Fe{\sc ii}. The results suggest an upper atmospheric temperature ($\sim15000$\,K), higher than that predicted by current state-of-the-art hydrodynamic models.
\end{abstract}


\section{Introduction} \label{sec:intro}
WASP-189b is an UHJ orbiting a bright ($V$\,=\,6.64\,mag) A4 star with a period of about 2.7\,days \citep{Anderson2018,Lendl202}. UHJs are planets with equilibrium temperatures ($T_{\rm eq}$) higher than 2000\,K and for which the continuum absorption level in the optical and near-infrared is mainly controlled by H$^-$ opacity \citep[e.g.][]{arcangeli2018,parmentier2018,Fossati2021}. The high $T_{\rm eq}$, and thus large pressure scale height, of these planets favor atmospheric characterization observations both in transmission and emission. In the case of WASP-189b, \citet{Lendl202} collected high-precision optical primary transit and secondary eclipse photometry with the CHaracterising ExOPlanets Satellite (CHEOPS) satellite \citep{Benz2021} and inferred that the day-side planetary atmosphere is rather unreflective (low albedo; \cite{Lendl202}) at a temperature of about 3400\,K, when assuming inefficient heat redistribution. Recently, \citet{Prinoth2022} collected and analysed HARPS (High Accuracy Radial Velocity Planet Searcher) high-resolution optical spectra of WASP-189b during primary transit, which led them to detect absorption by TiO and several metals such as Fe, Ti, Cr, Mg, V, and Mn. Their analysis found absorption depths attributed to neutral atoms, consistent with local thermodynamic equilibrium (LTE) based on hydrostatic and chemical equilibrium models. They also noted a deviation of their model from observations for strong metal ions and suggested non-LTE effects, hydrodynamic escape or night-side condensation as possible causes.

UHJs are typically found orbiting bright intermediate-mass stars \citep{Casasayas2019}, with a large range of high-energy (X-ray and extreme ultraviolet, EUV; together XUV) luminosity ranging from a few times solar (later than spectral type A3/A4) to orders of magnitude sub-solar \citep[earlier than spectral type A3/A4;][]{fossati2018b}. The high atmospheric temperatures of UHJs and the fact that they typically orbit early type stars (F- and A-type) that can have very strong (later than spectral type A4) or very weak (earlier than spectral type A4) XUV emission make these planets ideal laboratories for studying the physics of planetary upper atmospheres and, in particular, the separate roles of XUV and UV irradiation in heating the atmosphere and driving escape \citep{fossati2018b}.

Atmospheric escape from exoplanets can be observed with excited metal, He, and H lines (typically H Lyman $\alpha$, H Balmer lines and He{\sc i}\,(2$^3$S) triplet at $\approx$10830\,\AA\ in addition to neutral and ionized metals). Ultraviolet (UV) wavelengths probe many of these lines \citep[see e.g.][]{fossati2015}\footnote{The metastable He{\sc i}\,(2$^3$S) triplet at $\approx$10830\,\AA\ can probe upper atmospheres in alternative to the ultraviolet (UV) band, but the spectral energy distribution (SED) of intermediate-mass stars inhibits the formation of He{\sc i}\,(2$^3$S) \citep{oklopcic2018,oklopcic2019}.}. UV transmission spectroscopy observations are typically conducted at far-UV (FUV) wavelengths (e.g. Ly$\alpha$), but in the near-UV (NUV) band stars have a significantly stronger and more uniform emission than in the FUV \citep{haswell2012,llama2015}. Furthermore, the NUV range contains strong lines of several abundant metals, such as Mg and Fe, that effectively probe upper atmospheres and escape \citep{fossati10,haswell2012,SingEtal2019ajWASP121bTransmissionNUV,CubillosEtal2020ajHD209458bNUV,Cubillos2023}, as well as molecular bands probing atmospheric haze and cloud condensations \citep{Lothringer2022}.

We present the analysis and results obtained from three NUV transit observations of WASP-189b collected with the Colorado Ultraviolet Transit Experiment ($CUTE$) mission, which is a NASA 6U CubeSat (with stowed dimensions of $11.0\times23.7\times36.2~cm$) CubeSat carrying onboard a rectangular Cassegrain telescope feeding light into a low-resolution NUV (2479~--~3306 \AA) spectrograph \citep{FlemingEtal2018jatisCUTE,kevin2022, Egan2022}. $CUTE$ monitors transiting exoplanets orbiting bright stars to study the upper atmospheres and star-planet interaction processes. The early spectral type and brightness of the host star make the WASP-189 system ideal for $CUTE$ observations \citep{sreejith2019}.

This paper is organised as follows. We describe the observations in Section \ref{sec:obs} and the data analysis in Section \ref{sec:da}. We discuss the results and their implications in Section \ref{sec:results}. Finally, in Section \ref{sec:summary} we summarises our findings and future work based on $CUTE$ data.

\section{Observations} \label{sec:obs}

We observed WASP-189b for three transits with $CUTE$ on 2022-03-23, 2022-03-26, and 2022-03-28 (visits 1, 2, and 3, respectively). Each transit (one $CUTE$ visit) was observed over a period approximately five times the transit duration covering the planet's orbital phase from -0.2 to 0.2. A typical $CUTE$ orbit is 96 minutes and we observed the target for sixteen orbits in visits two and three, while in visit one the target has been observed for just thirteen orbits due to observational constraints and a spacecraft reset \citep{kevin2022}. The number of exposures per satellite orbit varies from one to five, depending on target position and pointing constraints that include the Sun angle, the Moon angle, the elevation cut off with Earth's limb, the spacecraft settling time, and the location of the satellite with respect to south Atlantic anomaly and polar keep out zones \citep[][under review]{Ambily2023}. These constraints, combined with the spacecraft settling time of up to 15 minutes and data processing time of 70 seconds per exposure, limit the total number of exposures per eclipse to five.   
 Each 300 second $CUTE$ exposure consists of a two-dimensional spectrogram of 2048$\times$100 pixels centred around the spectrum covering the $2480-3306$ \AA\, spectral range with a spectral resolving power of about 1000 with a SNR of about 25 per resolution element for a 300 second exposure. The CCD is passively cooled and maintains the temperature between $-10$~to~$-5^{\circ}C$ with occasional spikes in temperature during ground station passes and pointing anomalies. $CUTE$, with its single star tracker, provides pointing jitter (standard deviation) of better than 6\arcsec~for about 60\% of the observations \citep{Egan2022}, but some observations after the orbital day side and close to the South Atlantic Anomaly (SAA) are known to have higher jitter, leading to lower resolution and noisier spectra. We refer readers to \citet{Egan2022} for a detailed discussion on $CUTE$'s on-orbit performance. The data were reduced following the methods described by \citet{sreejith2022} to obtain wavelength and flux-calibrated one-dimensional spectra. 

\section{Data Analysis} \label{sec:da}
Our data analysis scheme follows a similar methodology as described by \citet{CubillosEtal2020ajHD209458bNUV,Cubillos2023}. For each visit, we first construct ``white'' light curves by integrating the flux in each exposure over the wavelength range to increase our signal-to-noise ratio \citep{KreidbergEtal2014natCloudsGJ1214b}. This white light curve is used to characterize the instrumental systematics. Then, we divide the systematics model from the raw data and combine the systematics-corrected data from all visits into spectral light curves at a range of spectral binnings, and extract the transmission spectra of the target as explained in detail in section~\ref{sec:white}.

We carry out an initial data quality screening on our observations to remove outliers. We select all observations where the RA and Dec jitter values are below 6\arcsec~and the CCD temperature is below $-5^{\circ}C$. This cut-off is to enable uniformity in all the data points, as observations with larger jitter will have lower fluxes due to part of the spectrum falling outside our extraction region, and possibly vignetting of the telescope beam by the edges of the slit \citep{sreejith2022}. These selection criteria are applied to both visits 2 and 3.  In the case of visit 1 due to a spacecraft reset, RA and Dec jitter information was not available and we incorporated a flux cut-off to reject bad jitter observations, based on the findings from visits 2 and 3.  We also noticed that observations after an orbital day (Sun-lit side)  lead to low fluxes from visits 1 and visit 2 (in visit 3 these observations where also flagged as bad jitter) possibly due to scatter effects as in these latitudes we observe through a much larger column of the Earth's atmosphere. 


\subsection{Individual visits}\label{sec:white}
We identify systematics using a visit-specific white light analysis, integrating the entire spectrum to create each individual data point. This  assumes that the instrumental systematics behave similarly over the integrated range \citep{KreidbergEtal2014natCloudsGJ1214b}. We fit the raw light curves ($F_\lambda(t)$) with a parametric transit and systematics models similar to  \cite{Cubillos2023}
as a function of time ($t$), $CUTE$ orbital phase ($\phi$), and jitter vector ($j_i$):

\begin{equation}
F_\lambda(t) = T_{\lambda}(t)\;S_{\lambda}(t, \phi, j_i),
\end{equation}
where $T_{\lambda}(t)$ is a
\citet{MandelAgol2002apjLightcurves} transit model and $S_{\lambda}(t, \phi,j_i)$ is a model of the $CUTE$ instrumental systematics similar to the scheme used in \cite{Cubillos2023}. We obtain the best-fitting parameters and uncertainties from a Levenberg-Marquardt optimization and a Markov-chain Monte Carlo (MCMC) sampling, respectively, employing the open-source \textsc{mc3} package\footnote{\href{https://mc3.readthedocs.io/}
{https://mc3.readthedocs.io/}} \citep{CubillosEtal2017apjRednoise}, which implements the Snooker Differential-evolution MCMC algorithm of \citet[][]{terBraak2008SnookerDEMC}. This enables us to arrive at statistically robust parameter estimations.

Our transit model fits the input data using the planet orbital parameters, planet-star radius ratios $R_p/R_{s}(\lambda$), out-of-transit stellar fluxes $F\sb{\rm s}(\lambda)$, and limb-darkening coefficients.  We used the planet orbital parameters computed by \cite{Lendl202} using CHEOPS observations. These parameters, except for mid-transit time, $R_p/R_{s}(\lambda$), and $F\sb{\rm s}(\lambda)$, were kept fixed during the fitting and MCMC process. The mid-transit time is fitted by applying a Gaussian prior according to the measured value from \cite{Lendl202}. Uncertainties are propagated  according to the epoch of the $CUTE$ observations. The limb-darkening parameters were computed based on the stellar properties using the open-source code of \citet{EspinozaJordan2015mnrasLimbDarkeningI}. This enables us to compute the limb-darkening coefficients over our wavelength range based on the PHOENIX stellar model \citep{HusserEtal2013aaPHOENIXstellarModels} according to WASP-189 stellar properties from \citet{Lendl202}.

The effects of systematics on transit measurements are well documented in the literature \citep[e.g.][]{BrownEtal2001apjHD209458bHSTstis,WakefordEtal2016apjHSTsystematics,SingEtal2019ajWASP121bTransmissionNUV, CubillosEtal2020ajHD209458bNUV, Cubillos2023}. Similar to the systematics accounted for in HST/STIS observations, we use a time-dependent polynomial that can account for stellar activity and another polynomial that can account for orbit-dependent systematics. Building on the work of  \cite{SingEtal2019ajWASP121bTransmissionNUV} and \cite{Cubillos2023}, we also make use of jitter data to decorrelate the instrumental systematics from the transit signal. The jitter parameters we use are summarised in Table~\ref{table:table2}. Employing these three sets of information we have a polynomial systematic model that is up to cubic in time, quartic in telescope phase and linear in one of the jitter parameters as shown below:

\begin{align}
\nonumber
& S_{\lambda}(t, \phi, j_i) =  1 +\ a_{0}(t-t_{0}) + a_{1}(t-t_{0})^{2} + a_{2}(t-t_{0})^{3}  \\
\nonumber
& \hspace{1.0cm} + b_{0}(\phi-\phi_{0}) + b_{1}(\phi-\phi_{0})^{2} 
  +\ b_{2}(\phi-\phi_{0})^{3} + b_{3}(\phi-\phi_{0})^{4} \\
& \hspace{1.0cm} + c_0(j_i-\langle{j_i}\rangle),
\label{eq:systematics}
\end{align}
where $a_{k}$, $b_{k}$, and $c_{k}$ are the polynomial coefficients of the fit and $t\sb{0}$ and $\phi\sb{0}$ are the reference values for the time and phase, set as mid-transit time $t\sb{0}=T\sb{0}$ and the $CUTE$ telescope mid-phase $\phi\sb{0}=0.2$, respectively. $\langle{j_i}\rangle$ denotes the mean
value of the jitter parameter $j_i$ along the visit as summarised in Table \ref{table:table2}. 

\begin{table}
\begin{center}
\begin{tabular}{l l }          
\hline\hline                        
Jitter parameters &  \\
\hline
CCD temperature (TMP) & Median of CCD temperature during a 300 second exposure. \\
RA jitter (RAJ) & Standard deviation of RA during a 300 second exposure.\\
Dec jitter  (DCJ )& Standard deviation of Dec during a 300 second exposure.\\
Roll jitter  (ROJ) & Standard deviation of Roll during a 300 second exposure. \\ 
Geo latitude (GLA) & Latitude of the satellite during a 300 second exposure.\\
Geo longitude (GLO) & Longitude of the satellite during a 300 second exposure. \\
Geo  altitude (GAL) & Altitude of the satellite during a 300 second exposure. \\
Sun angle  (SAN) & Angle between the telescope boresight and the Sun \\
\hline
\end{tabular}
\caption{\label{table:table2}Jitter parameters}
\end{center}
\end{table}

We make use of both Bayesian model selection approach
\citep[see, e.g.,][]{Trotta2007mnrasBayesianModelSelection} and Akaike Information Criterion corrected for small sample sizes \citep[AICc,][]{Liddle2007mnrasBIC} for model selection. All combinations of polynomials and jitter parameters from equation~\ref{eq:systematics} are compared for this approach. Both approaches in the case of our analysis prefer models which minimizes both the Bayesian Information Criterion (BIC) and AIC values. We refer the reader to \cite{Cubillos2023} for an additional general discussion of BIC and AIC comparison used in our analysis methods. 

\begin{table}[t]
\centering
\begin{tabular} {@{\extracolsep{-0.1cm}} lccccc}
\hline
\hline
Systematics model & BIC   & AICc         & $\chi^2_{\rm red}$ & $R_p$/$R_s$ & Phase offset\\
{($t$-deg, $\phi$-deg, jitter)} &  &  &   \\
\hline
\multicolumn{1}{l}{\bf Visit 1} \\
(3, 2, GAL)      & 46.54   &  46.47  &  1.01 & $0.101_{-0.016}^{+0.009}$ & $0.0148_{-0.005}^{+0.003}$\\
\multicolumn{1}{l}{\bf Visit 2} \\
(1, 4, GLA) &  113.83  &  108.19 &   3.4319 & $0.107_{-0.020}^{+0.006}$ & $0.0162_{-0.0057}^{+0.0492}$ \\
\multicolumn{1}{l}{\bf Visit 3} \\
(1, 2, GLA) & 71.10  &   65.11 &   1.7933 & $0.108_{-0.014}^{+0.010}$ & $-0.0016_{-0.0028}^{+0.0073}$\\
\hline
\end{tabular}
\caption{BIC/AICc best-fit Models. The polynomial degree and jitter parameter of systematics model is shown in the first column.}
\label{table:statistics}
\end{table}

\begin{figure*}
\centering
\includegraphics[width=\textwidth]{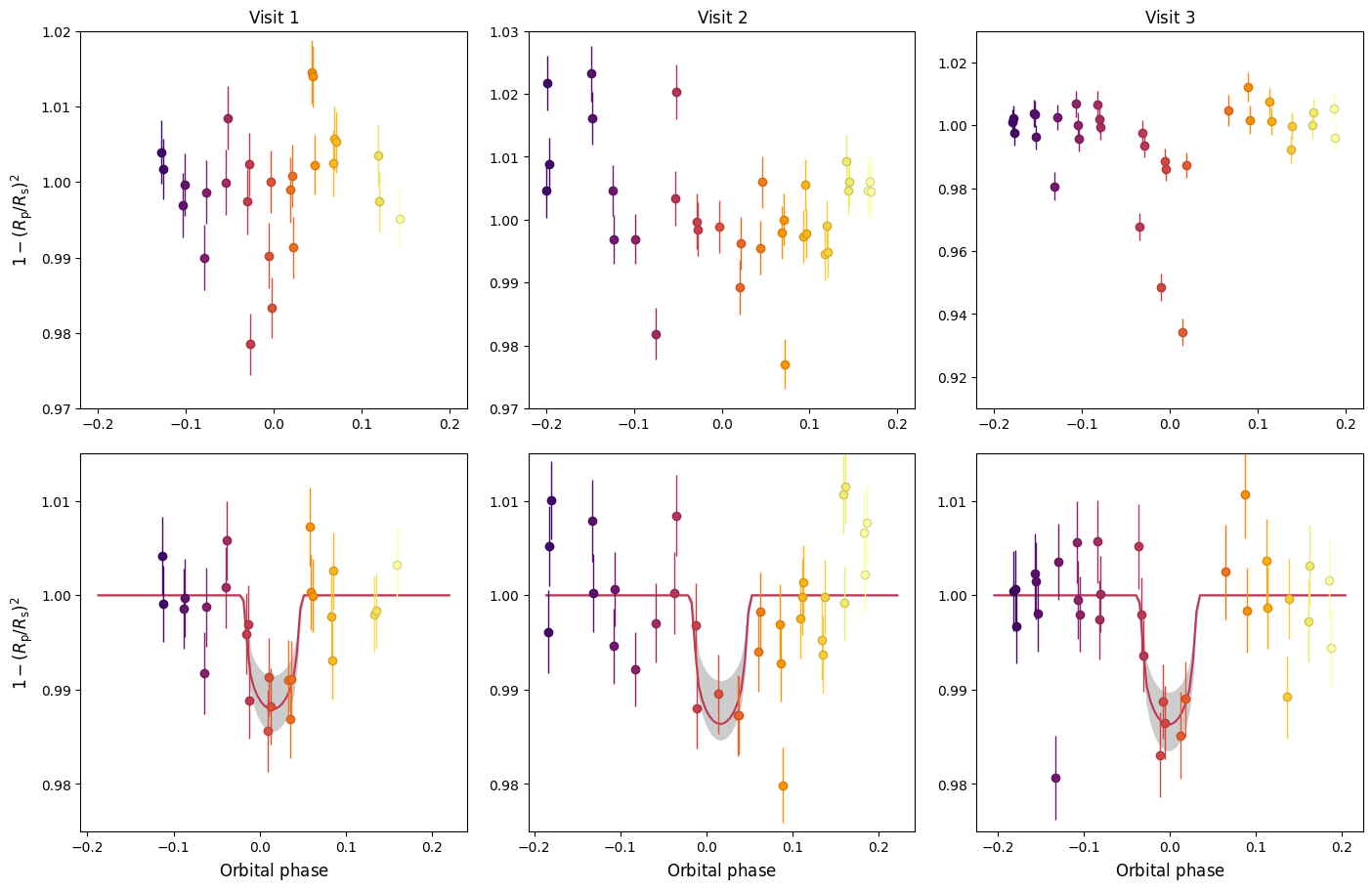}
\caption{ NUV White light-curve fitting of WASP-189b.
  The top panels shows the raw light-curve. The 
  bottom panels show the systematics-corrected light curves following the best-BIC/best-AICc.
    The solid orange lines denote the transit model of the best-fit. The data points are colored based on the $CUTE$ satellite orbits (from purple to yellow in increasing order).}
\label{fig:white_light_curves}
\end{figure*}

Table \ref{table:statistics} summarises the best-fitting models for each visit. We see visit-specific and $CUTE$ orbit-specific systematics in our data. Our analysis indicates stronger systematics in visit 2 compared to the other two visits. This is evident by the larger BIC, AIC and reduced $\chi^{2}$ values. We found that the fits were optimized with a phase offset between the UV and optical light curves of about $\sim$0.015, although it is unclear if this is a physical effect or an artifact of the CUTE data collection systematics. Our analysis gives self-consistent transit depths for all visits as shown in Figure~\ref{fig:white_light_curves} with best-fit polynomials in time and telescope phase within the range found for lightcurve detrending in  HST STIS analysis \citep{Cubillos2023,Gressier2023}. 


\subsubsection{Divide-white Spectral Extraction}

The $CUTE$ transmission spectra are obtained by the ``divide-white'' spectral analysis method of \citet{KreidbergEtal2014natCloudsGJ1214b}, described in detail for NUV transit spectroscopy in {\citet{CubillosEtal2020ajHD209458bNUV}}. We create visit-specific non-parametric instrumental systematics by dividing the white-light curves by the best-fit transit model (see Sect.~\ref{sec:white}). This is based on the assumption that the instrumental systematics have a weak wavelength dependence, guided by our analysis of raw light curves. These systematics are divided out of the raw spectral light curve and the resultant light curve are dominated by the transit signal where instrument systematics  have been mitigated.

We estimate uncertainties of the best-fit white light transit model from the standard deviation of the distribution for the white-light transit model, generated from the white-analysis posterior distribution. Then we use the error-propagation formula \citep{Bevington&Robinson2003} to account for all uncertainties throughout the steps involved to construct the non-parametric systematics model and obtain the systematics-corrected light curve.

This systematics-corrected data is divided into wavelength bins of our choice and co-added to create wavelength-specific data points. These data are fitted with a \citet{MandelAgol2002apjLightcurves} transit model with a fixed orbital period ($P$), inclination ($i$) and semi-major-axis to stellar radius ratio ($a/R_{s}$) parameters, and a Gaussian prior for the mid-transit epoch based on our results from the white-light fit. The limb-darkening coefficients are also calculated at each wavelength bin from the stellar model as described in section~\ref{sec:white} and are kept fixed during the fit. Hence the free parameters for these fits are the mid-transit time, the transit depth and the out-of-transit flux in each wavelength bin. Since the light curve combines data from different epochs, we fit the out-of-transit flux level of each visit as an individual free parameter.

\begin{figure*}
\centering
\includegraphics[width=0.95\textwidth]{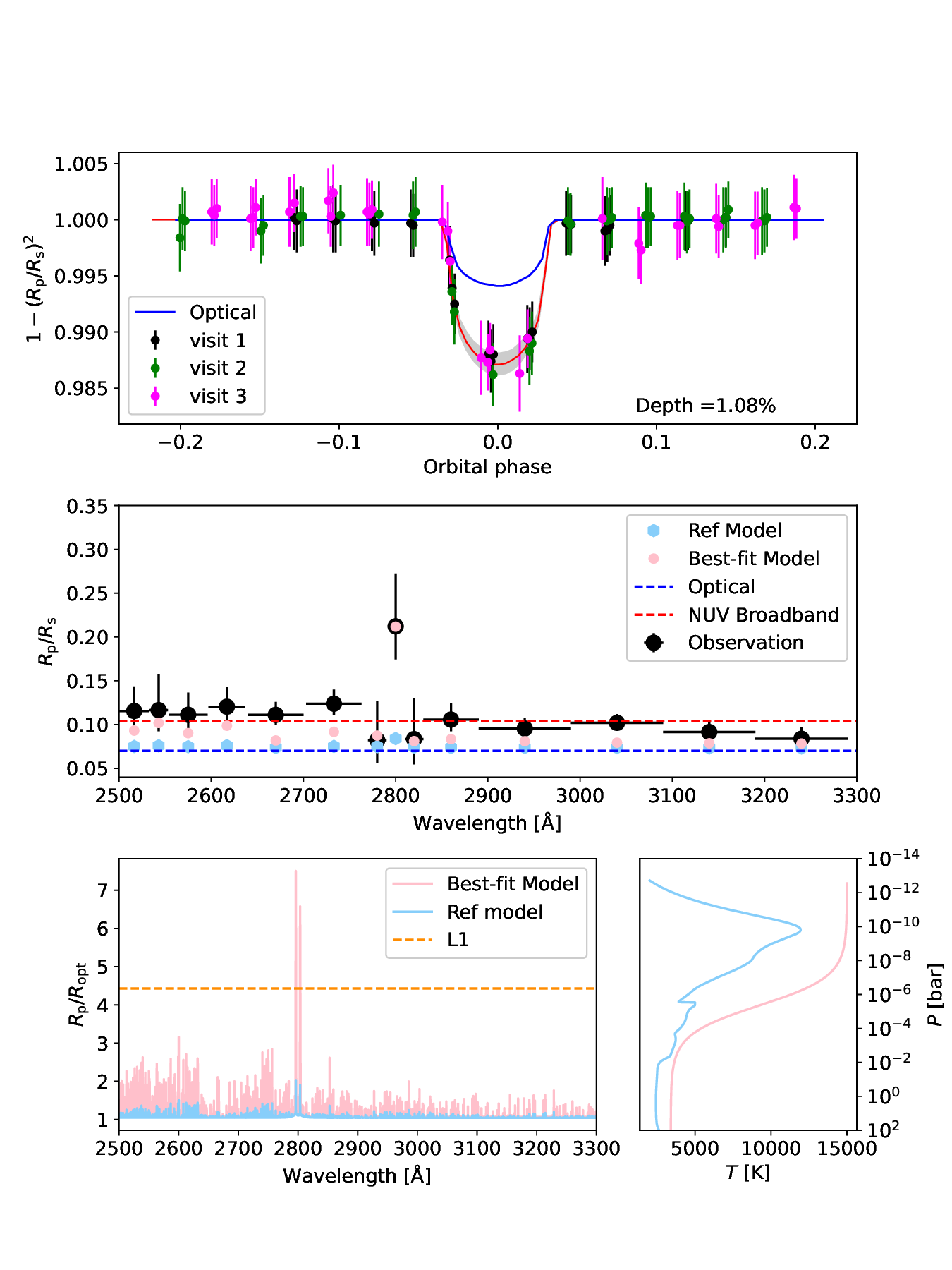}
\caption{ NUV Broadband light curve and transmission spectrum obtained from WASP-189b analysis. {\bf Top}: Broadband NUV light curve of WASP-189b for a 800~\AA~bin centered at 2900~\AA. {\bf Middle}: Transmission spectrum at a variable resolutions (black filled circles) with reference and best-fit models (light blue and pink respectively). {\bf Bottom-Left}: High-resolution models showing the variation of the ratio of planetary radius ($R_p$) with optical planetary radius ($R_{opt}$) as a function of wavelength. {\bf Bottom-Right}: T-P profiles used to generate the transmission spectrum (reference model in light blue and the best-fit model in pink).}
\label{fig:TS_both}
\end{figure*}

\section{Results and Discussion} \label{sec:results}
Figure \ref{fig:TS_both} (Top) shows the combined transit light curve in an  800~\AA~bin centred at 2900~\AA~(full $CUTE$ band). This light curve is generated using the divide-white method described in the previous section and points towards a transit depth of $1.08^{+0.08}_{-0.08}\%$, larger than the optical transit depth \citep[$0.496\%$;][]{Lendl202} of WASP-189b and consistent with the previous HST observations of UHJs \citep{SingEtal2019ajWASP121bTransmissionNUV,CubillosEtal2020ajHD209458bNUV,Lothringer2022}. Broadband NUV transits with this depth suggest we are probing the upper atmosphere of this planet that is unaffected by clouds and hazes and hence the broadband transit we observe would most likely be due to a combination of metal lines in the upper atmosphere of the planet. To understand the origin of this excess absorption we create a transmission spectrum at finer wavelength bins as shown in Figure \ref{fig:TS_both}. The $R_p/R_{s}(\lambda$) ratio increases as we move towards shorter wavelengths as observed for WASP-121b and WASP-178b \citep{SingEtal2019ajWASP121bTransmissionNUV,Lothringer2022}. This indicates stronger NUV absorption by WASP-189b at the shortest NUV wavelengths sampled by CUTE (2500-2800 \AA), with an NUV planet radius ($R_p/R_{s}$) of about $0.127_{-0.010}^{+0.012}$. Figure~\ref{fig:TS_both} (middle) displays the transmission spectrum in different wavelength bins. This transmission spectrum indicates significant evidence of planetary absorption of the Mg{\sc ii} h\&k lines (2802.705\,\AA\ and 2795.528\,\AA)\,. The absorption peak at 2800\,\AA\, is consistent with a model spectrum with strong Mg{\sc ii} h\&k lines, as shown in Figure~\ref{fig:TS_both} (bottom-left) and discussed below. The 10~\AA~bin centred at 2800~\AA\, has a planetary radius ratio ($R_p/R_{s}$) of $0.212^{+0.038}_{-0.061}$, which is significantly higher than the broadband NUV radius ratio of 0.105.  This is $\approx$3 times the optical radius of the planet at a significance of $\sim4.15\sigma$. 
This analysis indicate Mg{\sc ii} ions beyond the L1 Lagrange point for this planet and hence escaping. We provide our quantitative analysis of the mass-loss rate in the following section. 


To constrain the extent of the atmosphere of WASP-189b based on the observed NUV transit depths, we first construct a physical model of the whole atmosphere to predict density profiles of the possible absorbers (hereafter, the reference model). We model the lower and middle atmosphere with the photochemical-thermal structure model of \cite{Lavvas2021}, assuming global redistribution of heat and constituents. At pressures lower than 1 $\mu$bar, we use the multi-species hydrodynamic escape model of \cite{KoskinenEtal2022apjExtremePlanetsLoss}, updated to include the elements H, He, Mg, Si, Fe, O, C, N, Na, K, S, Ca, and relevant ions \citep{huang2023}. For the stellar spectrum, ranging from the FUV to the infrared, we use the spectral energy distribution (SED) computed with the PHOENIX stellar atmosphere code \citep{HusserEtal2013aaPHOENIXstellarModels} 
 based on the stellar parameters from \cite{Lendl202}. In the X-ray and EUV range, we follow the results of \cite{fossati2018b} and multiply the solar emission spectrum from \cite{Claire2012} by a factor of 3. 

Figure~\ref{fig:TS_both} (bottom-right) shows the temperature-pressure (T-P) profile predicted by our reference model. The discontinuity at the lower boundary of the escape model indicates that the radiative transfer schemes of the two models are not yet perfectly self-consistent but the results are close enough for illustration purposes. A more refined coupling will be pursued in future work. Figure~\ref{fig:TS_both} (middle) also compares the transit depths based on the reference model with the observations, showing that the model falls well short of the observations. Two main reasons contribute to this discrepancy. First, the radiative cooling rates in the thermosphere due to H and metal line emissions are relatively high and offset much of the stellar XUV heating that would otherwise power escape. This lowers the temperature in the upper atmosphere and reduces the mass loss rate. Second, the predicted absorption in the NUV is mostly due to Fe and Mg ions and second ionization of these elements by stellar XUV radiation is significant, reducing the density of the first ionization states.

To arrive at our best-fit model, we follow a simpler, more empirical approach. Effectively, we fix the T-P profile in the model until the predicted upper atmospheric opacities match the observations. The best-fit temperature profile is also shown in Figure~\ref{fig:TS_both}. We increase the temperature in the lower and middle atmosphere to the day-side temperature of 3400 K \citep{Lendl202}. We now model the composition in the lower and middle atmosphere by using the chemical equilibrium model GGChem \citep{GGchem2018}. This is acceptable because the temperature in the middle atmosphere is so high that Fe and Mg are strongly ionized thermally. In the upper atmosphere, we disable the energy equation solver and run the escape model with a fixed temperature profile. This allows us to still include photoionization of the metal ions at high altitudes. Our best-fit model is designed to match the Mg{\sc ii} transit depth, as illustrated in Figure~\ref{fig:TS_both} (bottom-left). Intriguingly, the shape of the model transit depths is also close to the observations outside of the Mg{\sc ii} doublet, particularly on the blue side, indicating additional opacity from Fe{\sc ii} absorption lines.

The best-fit model implies a mass loss rate of about \SI{4e8}{\kg\per\second}, which is more than 300 times higher than the mass loss rate predicted by our reference model. This new mass loss rate, however, is still consistent with stellar XUV energy-limited mass loss rate \citep[e.g.,][]{Erkaev2007} with a heating efficiency of about 10\% in the upper atmosphere (assuming that escape is powered by stellar radiation at 0.1-100 nm, which is conservative in this case). The best-fit temperature profile is also significantly hotter than the reference model temperature profile in the upper atmosphere. This could be explained either by an additional source of direct heating or lower radiative cooling rates. The difference between the observed transit depths that coincide with the Fe{\sc ii} lines and the best-fit model could arise from uncertainties in photoionization and recombination rates. A higher fraction of Fe{\sc ii} over Fe{\sc iii} could produce a larger transit depth that is still consistent with solar abundances.

\section{Summary and Conclusions} \label{sec:summary}

We present the first near-ultraviolet transit observations of WASP-189 from the $CUTE$ CubeSat. This is one of the hottest planets that has so far been observed in the NUV. Our findings of excess NUV absorption in the planet are consistent with previous HST transit observations of similar exoplanets \citep{SingEtal2019ajWASP121bTransmissionNUV,fossati10}. We find a dramatic increase in absorption below 3000 \AA, with transit depths decreasing towards the optical value at $\lambda$~$>$~3100~\AA. The Mg{\sc ii} h\&k lines are observed at a depth of about 3 times the visible planetary transit depth in a 10~\AA~bin centered on the Mg{\sc ii} h\&k lines at 2800\AA. Our modeling indicates the presence of of Mg{\sc ii} and possibly Fe{\sc ii} ions in the upper atmosphere. A fit to the Mg{\sc ii} transit depth indicates that Mg ions are escaping the atmosphere. Based on our best fit model to these observations that constrain the extent and temperature of the upper atmosphere, the total mass loss rate of all species is \SI{4e8}{\kg\per\second}. We attempted to reproduce the observed transmission spectrum with a state-of-the-art hydrodynamic upper atmosphere model accounting for the most relevant known atmospheric processes, but this led to an underestimate of the transit depths. Therefore, we increased the upper atmospheric temperature to fit the observed strength of the Mg{\sc ii} h\&k absorption. We found that the observations can be fitted with a temperature of about 15000\,K in the upper atmosphere, which is about 5000\,K higher than the peak temperature predicted by the self-consistent model.

There are just a few exoplanets with published NUV observations, and most have been spectroscopic observations obtained by HST \citep{fossati10,haswell2012,SingEtal2019ajWASP121bTransmissionNUV,wakeford2020,Lothringer2022, Cubillos2023, Gressier2023} and a few are photometric with multiple instruments \citep{Folsom2018,salz2019,Corrales2021}. Despite the differences between the systems observed in the NUV, an extended upper atmosphere seems to be a common feature. Some of the UHJs previously observed in the NUV orbit rather faint stars (e.g. WASP-12b, WASP-121b). Therefore, the host star brightness make WASP-189b a key target for future multi-wavelength observations aimed at deepening our understanding of these ultra-hot worlds.

\begin{acknowledgments}
{\bf Acknowledgments:} $CUTE$ is supported by NASA grant NNX17AI84G and 80NSSC21K1667 (PI - K. France) to the University of Colorado Boulder.
This project was funded in part by the Austrian Science Fund (FWF) J4595-N and J4596-N. L.F. acknowledge financial support from the Austrian Forschungsf\"orderungsgesellschaft FFG project CARNIVALS P885348. AV acknowledges funding from the European Research Council (ERC) under the European Union's Horizon 2020 research and innovation programme (grant agreement No 817540, ASTROFLOW). This research has made use of NASA's Astrophysics Data System Bibliographic Services.
\end{acknowledgments}
\vspace{5mm}
\facilities{$CUTE$}

\software{{\textsc{MC3}} \citep{CubillosEtal2017apjRednoise},
{\textsc{Numpy}} \citep{HarrisEtal2020natNumpy},
{\textsc{SciPy}} \citep{VirtanenEtal2020natmeScipy},
{\textsc{Matplotlib}} \citep{Hunter2007ieeeMatplotlib},
{\textsc{IPython}} \citep{PerezGranger2007cseIPython}, and
{\textsc{bibmanager}} \citep{Cubillos2019zndoBibmanager}.  }

\appendix
\section{Stellar and planetary parameters}
\label{sec:appendix1}
The stellar and planetary parameters used in the fit are as shown in Table~\ref{table:table_para}. 
\begin{table}
\begin{center}
\begin{tabular}{l c }          
\hline\hline                        
Orbital parameter &  \\
\hline
Orbital period (days) $P$ & $2.724033$ \\
inclination (degrees) $i$  & $84.03$\\
semi-major-axis to stellar radius ratio $a/R_{s}$  & $0.05053$ AU \\ 
Transit epoch (MJD) $T_0$ &     $58926.04169599991$\\
\hline
Stellar parameters & \\
\hline
Effective temperature (K)  & 8000 \\
log10(gravity) [cgs units] & 3.9 \\
Stellar metallicity [M/H]   & 0.29 \\
Microturbulent velocity (km/s) & 2.7 \\
\hline
\end{tabular}
\caption{WASP-189b orbital and stellar parameters input to calculate the best-fit $CUTE$ light curves.}
\label{table:table_para}
\end{center}
\end{table}

\section{transmission Spectrum of WASP-189b}
\label{sec:appendix2}
Table~\ref{table:hun} and~\ref{table:ten} represents truncated transmission spectrum at 100~\AA~and 10~\AA~ bin respectively. Full table will be made available online.
\begin{figure*}
\centering
\includegraphics[width=\textwidth]{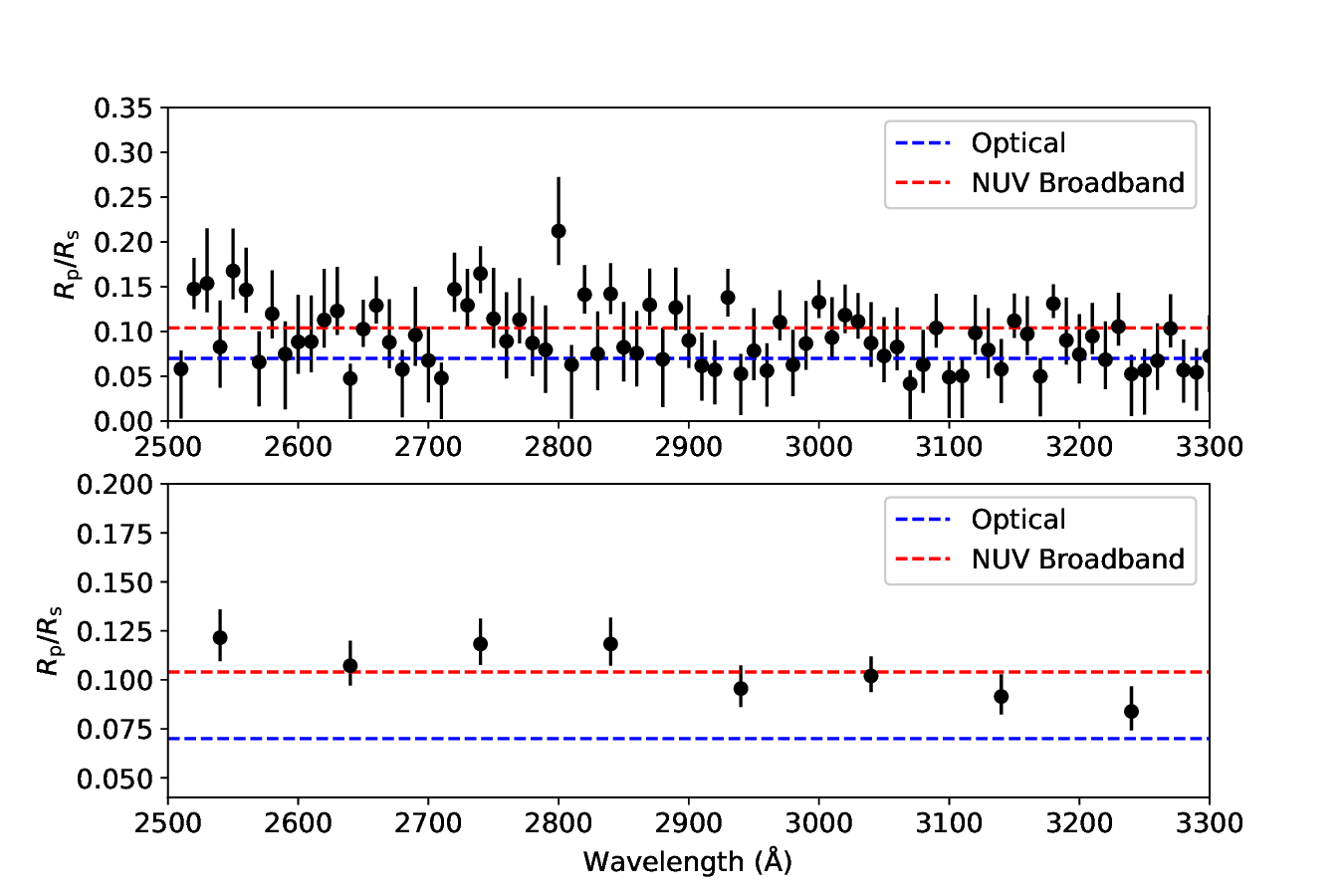}
\caption{{\bf Top}: Transmission spectrum at 10~\AA~ bin. {\bf Bottom}: Transmission spectrum at 100~\AA~bin.}
\label{fig:tp}
\end{figure*}

\begin{table}[h]
\centering
\begin{tabular} {lc}
\hline
\hline
$\lambda$ ($\AA$) & $R_p/R_{s}$ \\
\hline
$2540$  & $0.121_{-0.012}^{+0.015}$ \\
$2640$  & $0.107_{-0.010}^{+0.013}$ \\
$2740$  & $0.118_{-0.010}^{+0.013}$ \\
$2840$  & $0.118_{-0.011}^{+0.014}$ \\
$2940$  & $0.096_{-0.009}^{+0.012}$ \\
$3040$  & $0.102_{-0.008}^{+0.009}$ \\
$3140$  & $0.091_{-0.009}^{+0.011}$ \\
$3240$  & $0.084_{-0.009}^{+0.013}$ \\
\hline
\end{tabular}
\caption{ NUV Transmission Spectrum at 100~\AA~bin}
\label{table:hun}
\end{table}

\begin{table}[h]
\centering
\begin{tabular} {lc}
\hline
\hline
$\lambda$ ($\AA$) & $R_p/R_{s}$ \\
\hline
$2510$  & $0.0582_{-0.0554}^{+0.0207}$ \\
$2520$  & $0.1475_{-0.0223}^{+0.0345}$ \\
$2530$  & $0.1537_{-0.0322}^{+0.0614}$ \\
\hline
\end{tabular}
\caption{ NUV Transmission Spectrum at 10\AA~bin}
\label{table:ten}
\end{table}

\bibliography{wasp189}
\bibliographystyle{aasjournal}



\end{document}